  \documentclass[prl,aps,twocolumn,showpacs]{revtex4}
\usepackage[dvips]{graphicx}
\usepackage{dcolumn}%
\usepackage{amsmath}%
\usepackage{amsfonts}%
\usepackage{amssymb}
\providecommand{\U}[1]{\protect\rule{.1in}{.1in}}
\newcommand{\be}{\begin{equation}}
\newcommand{\ee}{\end{equation}}
\newcommand{\bea}{\begin{eqnarray}}
\newcommand{\eea}{\end{eqnarray}}
\newcommand{\nn}{ \nonumber}

\begin{document}

\title{Effect of Fermi-liquid interactions on the low-temperature de Haas-van Alphen oscillations in quasi-two-dimensional conductors}

\author{ Natalya A. Zimbovskaya}

\affiliation{Department of Physics and Electronics, University of Puerto Rico-Humacao, CUH Station, Humacao, PR 00791 \\and Institute for Functional Nanomaterials, University of Puerto Rico, San Juan, PR 00931}

\begin{abstract}
In this work we present the results of theoretical analysis of the de Haas-van
Alphen oscillations in quasi-two-dimensional conductors. We have been studying the effect of the Fermi-liquid correlations of charge carriers on the above oscillations. It was shown that at reasonably low temperatures and weak electron scattering the Fermi-liquid interactions may cause noticeable changes in both amplitude and shape of the oscillations even at realistically small values of the Fermi-liquid parameters. Also, we show that the Fermi-liquid interactions in the system of the charge carriers may cause magnetic instability of a quasi-two-dimensional conductor near the peaks of quantum oscillations in the electron density of states at the Fermi surface, indicating the possibility for the diamagnetic phase transition within the relevant ranges of the applied magnetic fields.
 \end{abstract}

\date{\today}
\maketitle



\section{I. Introduction}

Magnetic quantum oscillations \cite{1,2,3} have been recognized as one of the major tools
to map Fermi surfaces (FS) in metals. Analysis of the experimental data is now a well established procedure and is based on the classical paper by Lifshitz and Kosevich
(LK) \cite{4} who first layed out a quantitative theory of the de Haas-van
Alphen effect. The LK expression for the oscillating part of the thermodynamic
potential and magnetization of metallic electrons in a strong (quantizing)
magnetic field $\mathbf{B}$ was derived assuming that conduction electrons are
noninteracting quasiparticles in a periodic crystal potential. This potential
determines the electron dispersion, $E\mathbf{(p)\ (p}$ being the electron
quasimomentum), and therefore, the effective mass of conduction electrons
$m^{\ast}$ and their FS. In fact, conduction electrons interact with each
other. Studies of modifications of the LK results arising due to
electron-electron interactions within the general many-body quantum field
theoretical approach started in early sixties in the works of Luttinger
\cite{5}, and continued through the next three decades \cite{6}. It was shown
that electron-electron interactions may bring noticeable changes in the de
Haas-van Alphen oscillations which makes further analysis worthwhile.

One of the oldest and still powerful method to deal with electron-electron
interaction is the Landau Fermi-liquid (FL) theory \cite{7,8,9}. It is important to
realize that while the \textit{phenomenological} Fermi liquid theory and the
microscopic many-body perturbation theory (and, when applicable, the exact
density functional theory) by definition lead to the same observable
quantities, such as response to an external field, both zero approximation
and its renormalization depend on the taken approach. A discussion to this
effect in application to the dielectric response of metals can be found, for
instance, in Ref. \cite{10}. It is always instructive to look at the same
phenomenon from different points of view. Respecting the value of the
many-body perturbation theory as applied to quantum oscillations \cite{6}, we
emphasize that the Fermi-liquid theory has provided important insights in such
areas, relevant for the de Haas-van Alphen physics, as high frequency
collective modes in metals \cite{11,12,13,14,15,16}, or oscillations of various
thermodynamic observables in quantizing magnetic fields \cite{17,18,19,20}. 

An advantage of this phenomenological theory is that it enables to describe 
the effects of quasiparticles interactions in such a way that makes the
interpretation of the results rather transparent, as compared with the
field-theoretical methods. 
At the same time the many-body approach brings general but cumbersome results, and usually it takes great calculational efforts and/or significant simplifications to get suitable expressions for comparison with experimental data. Which is more important, adopted simplifications may lead to omission of some qualitative effects of electron-electron interactions in quantum oscillations, as we show below. 

In the last two decades an entire series of quasi-two-dimensional (Q2D) 
materials with metallic type conductivity has been synthesized. These are 
organic conductors belonging to the family of tetrathiafulvalene salts, 
dichalcogenides of transition metals, intercalated compounds and some other. 
At present, these materials attract a significant interest. Their electronic 
properties are intensively studied, and the de Haas-van Alphen effect is 
employed as a tool in these studies \cite{2,3}. Correspondingly, the theory 
of this effect in the Q2D materials is currently being developed  
\cite{21,22,23,24,25,26}. The present analysis of the effect of Fermi-liquid interactions on the de Haas-van Alphen oscillations contributes to the above theory. Also, the analysis is motivated by the special features in the electron spectra in Q2D materials providing better opportunities for the Fermi-liquid effects to be manifested, as was mentioned in some earlier works \cite{17,20}.

In this paper we  show how  renormalizations of conduction electron
characteristics arising from FL interactions affect quantum
oscillations, and we express them in the form appropriate for comparison with
experiments. Also, we show that a magnetic phase transition leading to
emergence of diamagnetic domains may happen at low temperatures when the
cyclotron quantum $\hbar\omega$ is very large compared to the temperature
 expressed in the energy units: $(\hbar\omega\gg k_{B}T)$. 
 We emphasize that the analyzed effects are different from the many-body renormalizations of the band structure. Within the phenomenological theory the latter are already included in the ground state of conduction electrons.

\section{ii. Fermi-liquid renormalizations of the conduction electron
characteristics}

Within the phenomenological Landau FL theory single quasiparticle energies get
renormalized, and the renormalization is determined by the distribution of
excited quasiparticles. Accordingly, the energy of a \textquotedblleft
bare\textquotedblright\ (noninteracting) quasiparticle $E_{0}\mathbf{(p)}$
moving in the effective crystal potential is replaced by the renormalized
energy defined by the relation \cite{7}:
 \begin{equation}
E_{\sigma}(\mathbf{p,r,}t)=E_{0}(\mathbf{p})+\sum_{\mathbf{p^{\prime}%
\sigma^{\prime}}}F(\mathbf{p,s;p^{\prime},s^{\prime}})\delta\rho
(\mathbf{p^{\prime},s^{\prime},r,}t).\label{1}%
 \end{equation}
 Note that $E_{0}\mathbf{(p)}$ here is the energy spectrum in the absence of any excited quasiparticles, that is, at equilibrium and at zero temperature, while $\delta\rho(\mathbf{p^{\prime},s^{\prime},r,}t)$ represents the nonequilibrium part of the electrons distribution function, which may depend
on both position $\mathbf{r}$ of the quasiparticle and time $t.$ Also,
$\mathbf{s}$, $\mathbf{s^{\prime}}$ are  spin Pauli matrices, $(\sigma$ is
the spin quantum number), and $F(\mathbf{p,s;p^{\prime},s^{\prime}})$ is the
Fermi-liquid kernel (Landau correlation function), which describes additional
renormalization of the quasiparticle spectrum due to interaction with other
excited quasiparticles (but not with all electrons in the system, which is
included in $E_{0}(\mathbf{p})$). 
Neglecting spin-orbit interactions, the Landau correlation function may be
written as:
 \be
F(\mathbf{p,s;p^{\prime},s^{\prime}})=\varphi( \mathbf{p,p^{\prime}}%
)+4\psi\mathbf{(p,p^{\prime})(\mathbf{ss^{\prime}).}}\label{2}%
 \ee

As follows from the Eq. \ref{1}, the conduction electron velocity
$\mathbf{v}=\nabla_{\mathbf{p}}E$ differs from the bare velocity
$\mathbf{v}_{0}=\nabla_{\mathbf{p}}E_{0}.$ To proceed in our analysis we need
to bring velocities $\mathbf{v }$ and $\mathbf{v}_{0}$ into correlation. For
brevity we do not explicitly write out the variables $\mathbf{r,}t$ in the Eq.
\ref{1} in  further  calculations. This omission does not influence the
results. Differentiating the Eq. \ref{1} we obtain:%
 \begin{equation}
\mathbf{v_{\sigma}(p) = v}_{0} \mathbf{(p)} + \nabla_{\mathbf{p}} \left[
\sum_{\mathbf{p^{\prime}\sigma^{\prime}}} F \mathbf{(p,s;p^{\prime},s^{\prime
})\delta\rho(p^{\prime},s^{\prime})}\right] .\label{3}%
  \end{equation}
 The electron distribution function $\rho\mathbf{(p,s})$ is the sum of the
equilibrium part $\rho_{0}$ (the latter coincides with the Fermi distrubution
function for quasiparticles with single particle energies  $E_{0}%
\mathbf{(p))}$ and the nonequlibrium correction $\delta\rho. $ Multiplying
both parts of the Eq. \ref{3} by $\rho\mathbf{(p,s) }$ and performing
summation over $\mathbf{p,\sigma}$ we get:
  \begin{align}
 \sum_{\bf p \sigma} & \rho\bf{(p,s)} \bf{v_{\sigma}(p)}  =
\bf{\sum_{p \sigma} \rho(p,s)v_0 (p)} 
  +     \bf {\sum_{p \sigma} \rho(p,s) \nabla_{\bf p}  }
   \nn\\   & \times    
\left[ \sum_{\bf p'\sigma'}
 F \bf{(p,s;p',s')\delta \rho(p',s')}\right].
 \label{4}
     \end{align}
The second term on the right hand side of the Eq. \ref{4} could be converted to the form
 \begin{equation}
- \sum_{\mathbf{p\sigma}} \left[ \sum_{\mathbf{p^{\prime}\sigma^{\prime}}} F
\mathbf{(p,s;p^{\prime},s^{\prime}) \delta\rho(p^{\prime},s^{\prime})}\right]
\mathbf{\nabla_{p} \rho(p,s).}\label{5}%
   \end{equation}

Keeping only the terms linear in $\delta\rho$ (which is supposed to be small
compared to the equilibrium part of the distribution function), we may
approximate $\mathbf{\nabla_{p} \rho}$ as $\mathbf{v_{\sigma}(p)} 
\frac{\partial f_{\mathbf{p\sigma}}}{\partial E_{\bf p \sigma}}. $ Here, $f $ is the Fermi distribution function and the
quasiparticle energies $E_{\sigma}\mathbf{(p)}$  correspond to the local
equilibrium of the electron liquid. Also,  assuming that the FS of a
considered metal possesses a center of symmetry, we get
 \begin{align}
\sum_{\mathbf{p\sigma}} \rho\mathbf{(p,s) v_{\sigma}(p) = \sum_{p \sigma}
\delta\rho(p,s) v_{\sigma}(p)}, \nn\\
\sum_{\bf p\sigma} \rho{\bf(p,s) v}_{0}{\bf(p)} = \sum_{\bf p \sigma}
\delta\rho{\bf (p,s)  v}_{0} \bf (p). 
\label{6}%
 \end{align}
  Then, using the well known relation $F \mathbf{(p,s;p^{\prime},s^{\prime})} =
F \mathbf{(p^{\prime},s^{\prime};p,s) }$ and carrying out the replacement
$\mathbf{p,s  \rightleftharpoons p^{\prime},s^{\prime}}$ in the sums included
in the Eq. \ref{5}, we could rewrite Eq. \ref{4} as
  \begin{align}
 \sum_{\mathbf{p\sigma}} & \delta\rho\mathbf{(p,s)} \left[ \mathbf{v}_{\sigma}
\mathbf{(p)} + \sum_{\mathbf{p\sigma}} \frac{\partial f_{\mathbf{p^{\prime
}\sigma^{\prime}}}}{\partial E_{\bf p^{\prime} \sigma^{\prime}}}
F\mathbf{(p,s;p^{\prime},s^{\prime})} \mathbf{v_{\sigma^{\prime}} (p^{\prime})
}\right] \nonumber\\ & = 
 \sum_{\mathbf{p\sigma}} \delta\rho\mathbf{(p,s)} \mathbf{v}_{0} \mathbf{(p).}%
\label{7}%
   \end{align}
 Solving this for $\mathbf{v }$, we obtain
 \begin{equation}
\mathbf{v_{\sigma}(p)=v}_{0}\mathbf{(p)} - \sum_{\mathbf{p^{\prime}%
,\sigma^{\prime}}}\frac{\partial f_{\mathbf{p^{\prime}\sigma^{\prime}}}%
}{\partial E_{\bf p^{\prime} \sigma^{\prime}}}F(\mathbf{p, s;p^{\prime},s^{\prime
}})\bf v_{\sigma^{\prime}}\mathbf{(p^{\prime}).}\label{8}%
 \end{equation}

Exact expressions for the functions $\mathbf{\varphi(p,p^{\prime})}$ and
$\mathbf{\psi(p,p^{\prime})}$ are of course unknown. The simplest approximation
is to treat them as constants. This approximation is reasonable as long as the
interaction of quasiparticles (located at $\mathbf{r}$ and $\mathbf{r^{\prime
},}$ respectively), is extremely short range, so that the interaction can be
approximated as $V(\mathbf{r,r^{\prime}})=I\delta(\mathbf{r-r^{\prime}}).$
Using this approximation one captures some FL effects but in general
case it is not sufficient.

As a next step, one may expand the Fermi-liquid functions in the Eq. \ref{2}
in basis functions respecting the crystal symmetry, such as Allen's Fermi
surface harmonics \cite{27}:
   \begin{align}
\varphi(\mathbf{p,p^{\prime}}) =\sum\limits_{j=1}^{d}\sum\limits_{m=1}^{d_{j}%
}\varphi_{j}(p,p^{\prime})R_{jm}(\theta,\Phi)R_{jm}^{\ast}(\theta^{\prime
},\Phi^{\prime}),\nonumber\label{9}\\
\psi(\mathbf{p,p^{\prime}}) =\sum\limits_{j=1}^{d}\sum\limits_{m=1}^{d_{j}%
}\psi_{j}(p,p^{\prime})R_{jm}(\theta,\Phi)R_{jm}^{\ast}(\theta^{\prime}%
,\Phi^{\prime}).
  \end{align}
  Here, we introduce spherical coordinates for $\mathbf{p:\ p}=(p,\theta
,\Phi);\ d$ is the order of the point group; index $j$ labels irreducible
representations of the group; $d_{j}$ is the dimension of the $j$-th
irreducible representation; $\{R_{jm}(\theta,\Phi)\}$ is a basis of the $j$-th
irreducible representation including $d_{j}$ functions.

For an isotropic metal the spherical harmonics $Y_{jm}$ can be used as the
basis. Including orbital moments up to $j=2$ we
have, for a cubic symmetry (cubic harmonics): 
 \begin{align}  
  &\left(\begin{array}{l}
\varphi (\mathbf{p,p^{\prime }}) \\
\psi (\mathbf{p,p^{\prime }})%
\end{array}%
  \right)  
= \left(
\begin{array}{l}
\varphi _{0} \\
\psi _{0}%
\end{array}%
\right) +\left(
\begin{array}{l}
\varphi _{1} \\
\psi _{1}%
\end{array}%
\right) (p_{x}p_{x}^{\prime }+p_{y}p_{y}^{\prime }+p_{z}p_{z}^{\prime })
 \nonumber \\  
 & \quad + \left(
\begin{array}{l}
\varphi _{21} \\
\psi _{21}%
\end{array}
\right) (p_{z}p_{z}^{\prime }p_{x}p_{x}^{\prime }+p_{z}p_{z}^{\prime
}p_{y}p_{y}^{\prime }+p_{x}p_{x}^{\prime }p_{y}p_{y}^{\prime })
 \nonumber \\ 
 & \quad +  \left(
\begin{array}{l}
\varphi _{22} \\
\psi _{22}%
\end{array}%
\right) (p_{x}^{2}-p_{y}^{2})(p_{x}^{\prime 2}-p_{y}^{\prime 2})
\nn\\
 & \quad + \frac{1}{3}\left(
\begin{array}{l}
\varphi _{22} \\
\psi _{22}%
\end{array}%
\right) (2p_{z}^{2}-p_{x}^{2}-p_{y}^{2})(2p_{z}^{\prime 2}-p_{x}^{\prime
2}-p_{y}^{\prime 2}).  \label{10}
\end{align}%
 The coefficients $\varphi,\psi$ are material dependent constants. 
A common feature of Q2D metals is their layered structure with a pronounced 
anisotropy of the electrical conductivity. In such materials electron energy 
only weakly depends on the quasimomentum projection $ p = \bf p n $ on the 
normal $\bf n $ to the layers plane. In further consideration we assume
 $ {\bf n} = (0,0,1)$ and we neglect the asymmetries of the electron spectrum in the layers planes. Then the relevant Fermi surface is axially symmetrical. 

For systems with an axial symmetry  this expression (\ref{10}) needs to be correspondingly modified.  For instance, in the first order we have
 \begin{align}
\left(
\begin{array}
[c]{l}%
\varphi(\mathbf{p,p^{\prime}})\\
\psi(\mathbf{p,p^{\prime}})
\end{array}
\right)  &  = 
\left(
\begin{array}
[c]{l}%
\varphi_{0}\\
\psi_{0}%
\end{array}
\right)  +\left(
\begin{array}
[c]{l}%
\varphi_{10}\\
\psi_{10}%
\end{array}
\right)  p_{z}p_{z}^{\prime}
\nonumber\\  
+\left(
\begin{array}
[c]{l}%
\varphi_{11}\\
\psi_{11}%
\end{array} 
\right) & (p_{x}p_{x}^{\prime}+p_{y}p_{y}^{\prime}).\label{11}%
\end{align}
 This expression will be used from now on in the present paper.

When an external magnetic field $\mathbf{B}=(0,0,B)$ is applied the spin
degeneracy of the single electron energies is lifted, and we can write:
 \begin{equation}
E_{0\sigma}(\mathbf{p})=E_{0}(\mathbf{p})+\sigma g\beta_{0}B\equiv
E_{0}(\mathbf{p})+\Delta E_{0}\label{12}%
 \end{equation}
  where $E_{0}(\mathbf{p})$ does not depend on the electron spin, $g$ is the
electron Lande factor, and $\beta_{0}=e\hbar/2m_{0}c$ is the Bohr magneton
$(m_{0}$ is the free electron mass). The nonequilibrium correction to the
electron distribution function satisfies the equation \cite{7}:
 \begin{equation}
\delta\rho(\mathbf{p,s})=\overline{\delta\rho}(\mathbf{p, s})+\frac{\partial
f_{\mathbf{p\sigma}}}{\partial E_{\mathbf{p\sigma}}}\sum_{\bf p^{\prime}%
,\sigma^{\prime}}F(\mathbf{{p},s, \mathbf{p}^{\prime},s^{\prime}})\delta
\rho(\mathbf{{p}^{\prime},s^{\prime}})\label{13}%
\end{equation}
  where $\overline{\delta\rho}(\mathbf{p,s})$ describes the deviation of the
electron liquid from the state of local equilibrium. When the deviation arises
due to the effect of the applied magnetic field $\overline{\delta\rho
}=-(\partial f_{\mathbf{p\sigma}}/\partial E_{\mathbf{p\sigma}})g\beta\sigma
B.$ Substituting Eq. \ref{11} into Eq. \ref{13} and using the result in Eq.
\ref{1} we get:
  \begin{equation}
\Delta E=\Delta E_{0}-b^{\ast}\sigma g\beta_{0}B\equiv\frac{\sigma g\beta
_{0}B}{1+b_{0}}.\label{14}%
\end{equation}
 where $b^{\ast}=b_{0}/(1+b_{0}),\ $and $b_{0}$ is a dimensionless parameter
describing FL interactions of the conduction electrons, namely:
$b_{0}=-\nu_{0}(0)\psi_{0}$ where $\nu_{0}(0)$ is the density of states of
noninteracting conduction electrons on the Fermi surface in the absence of the
magnetic field. This is nothing but the standard Stoner renormalization of the
paramagnetic susceptibility. Note that here $\psi_{0}$ plays the role of the
Stoner parameter $I=\left\langle \delta^{2}E_{xc}/\delta\rho_{\uparrow}%
\delta\rho_{\downarrow}\right\rangle $ in the density functional theory, or of
the contact Coulomb interaction in the many-body theory.

The Luttinger theorem dictates the Fermi surface volume. Therefore, the radius and the cross-sectional areas of the Fermi sphere associated with an isotropic Fermi liquid remain unchanged due to quasiparticles interactions. In realistic metals whose conduction electrons form anisotropic Fermi liquids one may expect some minor changes in the FS geometry to appear. Such effects could be considered elsewhere. In the present work we neglect them. So, in further consideration we assume that FL interactions do not affect the FS geometry. Then the
cross sectional areas of the Fermi surface $A(p_z)$ cut out by the planes
perpendicular to the magnetic field $\mathbf{B}$ do not change when
electron-electron interactions are accounted for. However, the cyclotron
masses of conduction electrons undergo renormalization due to the
electron-electron interactions. The cyclotron mass is defined as:
   \begin{equation}
m_{\perp}=\frac{1}{2\pi}\frac{\partial A}{\partial E}\bigg|_{E=\mu}\equiv
\oint\frac{dl}{v_{\perp}}.\label{15}%
\end{equation}
  Here, $dl=\sqrt{dp_{x}^{2}+dp_{y}^{2}}$ is the element of length along the
cyclotron orbit in the quasimomentum space, $v_{\perp}=\sqrt{v_{x}^{2}%
+v_{y}^{2}},\ v_{\alpha}=\partial E/\partial p_{\alpha}\ (\alpha=x,y),$ and
$\mu$ is the chemical potential of conduction electrons.

 Substituting Eqs. \ref{11} into the Eq. \ref{8} we get $v_{\perp}=v_{\perp0}/(1+a_{1})$ where $v_{\perp0}=\sqrt{v_{x0}^{2}+v_{y0}^{2}},$ and $a_{1}$ is related to the FL parameter $\varphi_{11}$ as follows:
   \begin{equation}
-\nu_{0}(0)p_{0}^{2}\varphi_{11}/3=a_{1}\label{16}%
\end{equation}
  where $p_{0}$ is the maximum value of the longitudinal component of
quasimomentum.  So we get:
  \begin{equation}
m_{\perp}=m_{\perp0}(1+a_{1}),\label{17}%
  \end{equation}
$m_{\perp0}$ being the cyclotron mass of noninteracting quasiparticles. In the
case of isotropic electron system the cyclotron mass $m_{\perp0}$ coincides
with the crystalline effective mass $m^{\ast}$ Therefore, our result agrees
with the standard isotropic FL theory.

 Other quantities, such as the chemical potential of conduction electrons and their compressibility,  may experience different renormalizations, as well. The latter, for instance, is renormalized by a factor $1/(1+a_{0})=1/[1-\nu_{0}(0)\varphi_{0}]$, and the former by the factor $(1+a_{0})$ \cite{7}. So, the renormalized density of states $\nu(0)$ appears in the expressions for the electron compressibility and the velocity of sound in metals.

 The model of the extremely short range (contact) Coulomb interaction between quasiparticles is often employed, while applying the many-body theoretical approach to study de Haas--van Alphen effect (see e.g. Ref. \cite{6}). Within the phenomenological FL theory this model results in the approximation of the functions $\varphi \bf (p,p') $ and $\bf \psi (p,p')$ by constants $\varphi_0 $ and $\psi_0 ,$ respectively. Such approximation enables us to get the Stoner renormalization of the paramagnetic susceptibility, and electron compressibility  as shown above. However, it misses Fermi-liquid effects associated with the subsequent FL coefficients included in the Eqs. \ref{9}--\ref{11}, which could be significant in renormalizations of other parameters characterizing the charge carriers such as their cyclotron masses.

\section{iii. quantum oscillations of the longitudinal velocity of charge-carriers in q2d conductors}

 In further calculations we adopt the commonly used tight-binding approximation for the charge carriers spectrum in a quasi-two-dimensional metal. So, when a quantizing magnetic field is applied, the charge carriers energies may be written in the form:
    \be 
 E_0 (n, p_z,\sigma) = \hbar \omega \left(n +\frac{1}{2}\right ) 
+ \sigma \hbar \omega_0 - 2t \cos
 \left(\pi \frac{p_z}{p_0}\right).\label{18}
   \ee
  where $ \hbar \omega_0 $ is the spin splitting energy,  $ t $  is the interlayer transport integral, and 
$ p_0 = \pi \hbar/L $ where $ L $ is the interlayer distance. This expression 
(18) describes single particle energies of noninteracting quasiparticles.
 Now, the relation of matrix elements of renormalized $\bf v_{\nu\nu'} $ and bare ${\bf v}_{0\nu\nu'} $ velocities in accordance with Eq. \ref{8} takes on the form \cite{9}:
   \be
  {\bf v}_{\nu\nu'} = {\bf v}_{0\nu\nu'} - \sum_{\nu_1\nu_2} 
\frac{f_{\nu_1} - f_{\nu_2}}{E_{\nu_1} - E_{\nu_2}} F_{\nu\nu'}^{\nu_1\nu_2} {\bf v}_{\nu_1\nu_2}.  \label{19} 
     \ee
  Here, $E_\nu $ is the quasiparticle energy including the correction arising due to the FL interactions, and $ \nu = \{\alpha,\sigma\} $ is the set of quantum numbers of an electron in the magnetic field. The subset $\alpha $ includes the orbital numbers $ n, p_z $ and $x_0 $ (the latter labels the positions of the cyclotron orbits centers). Also, $ F_{\nu\nu'}^{\nu_1\nu_2} = \varphi_{\alpha\alpha'}^{\alpha_1\alpha_2} + 4 \psi_{\alpha\alpha'}^{\alpha_1\alpha_2} {\bf(s s}_1) $ are the matrix elements of the Fermi-liquid kernel.

For an axially symmetrical FS the off-diagonal matrix elements of the longitudinal velocity vanish and we obtain:   
 \be
v_{\nu\nu}=v_{0\nu\nu} - \sum_{\nu_{1}}\frac{df_{\nu_{1}}}{dE_{\nu_{1}}}%
F_{\nu\nu}^{\nu_{1}\nu_{1}}v_{\nu_{1}\nu_{1}}\label{20}%
  \ee
    Substituting the expression for the Fermi-liquid kernel into the Eq. \ref{20}, we get:
   \be
v_{\nu\nu}=v_{\alpha\alpha}\delta_{\sigma\sigma^{\prime}}+\sigma
v_{\alpha\alpha}^{s}\label{21}%
  \ee
 where both $v_{\alpha\alpha}$ and $v_{\alpha\alpha}^{s}$ only depend on $p_z,$
so in the further calculations we will use the notation $v_{\alpha\alpha
}\equiv v(p_z),\ v_{\alpha\alpha}^{s}\equiv v^{s}(p_z).$ These matrix elements
could be found from the system of equations that results from Eqs. \ref{20}
and \ref{21}:
  \begin{align}
v(p_z)   =&\Big [v_{0}(p_z)-\sum_{\alpha_{1}}\varphi_{\alpha\alpha}^{\alpha
_{1}\alpha_{1}}\big (v(p_{1z})\Gamma_{\alpha_{1}\alpha_{1}} 
+v^{s}(p_{1z})\Gamma_{\alpha_{1}\alpha_{1}}^{s}\big)\Big],
   \label{22}\\
v^{s}(p_z)  = &-\sum_{\alpha_{1}}\psi_{\alpha\alpha}^{\alpha_{1}\alpha_{1}%
}\big (v(p_{1z})\Gamma_{\alpha_{1}\alpha_{1}}^{s}+v^{s}(p_{1z})\Gamma
_{\alpha_{1}\alpha_{1}}\big).\label{23}%
\end{align}
   Here,
  \be
\Gamma_{\alpha_{1}\alpha_{1}}=\sum_{\sigma_{1}}\frac{df_{\alpha_{1}%
\sigma_{1}}}{dE_{\alpha_{1}\sigma_{1}}}, \qquad
 \Gamma_{\alpha_{1}\alpha_{1}}^{s}=\sum_{\sigma_{1}}\frac{df_{\alpha
_{1}\sigma_{1}}}{dE_{\alpha_{1}\sigma_{1}}}\sigma_{1}.\label{24}%
    \ee
  The de Haas-van Alphen oscillations are observed in magnetic fields when the
Landau levels spacing is small compared to the chemical potential of electrons
$(\hbar\omega\ll\mu).$ Under these conditions we may approximate the Fermi-liquid kernel by its expression in the absence of the magnetic fields (Eq. \ref{2}).  Using the above-described
approximation of the Fermi-liquid functions $\mathbf{\varphi(p,p^{\prime}%
)\ \psi(p,p^{\prime})}$, Eqs. \ref{11}, we can solve Eqs.
\ref{22}, \ref{23}. To this end, we need some averages over the Fermi surface,
namely:
  \begin{align}
R  = &-\sum_{\alpha}\Gamma_{\alpha\alpha}v_{0}(p_z)p_z =-\frac{1}{4\pi\hbar\lambda^{2}}
  \nonumber\\ & \times
\sum_{n,\sigma}\int\frac{df (E_{n,\sigma
}(p_z))}{dE_{n,\sigma}(p_z)}v_{0}(p_z)p_zdp_z.\label{25}%
  \end{align} 
  \be
    R'  =  - \sum_\alpha \Gamma_{\alpha\alpha}^s v_0 (p_z) p_z. \label{26}
   \ee
    The Fermi-liquid effects enter the system \ref{22} \ref{23} through the averages $ A, A', B, B' $ closely  related to the Fermi-liquid parameters:
   \be
A = - \varphi_{10} \sum_\alpha \Gamma_{\alpha \alpha} p_z^2, \qquad
 A' = - \varphi_{10} \sum_\alpha \Gamma_{\alpha \alpha}^s p_z^2. \label{27}
  \ee
  The expressions for $ B, B' $ could be obtained replacing $ \varphi_{10}$ 
by $ \psi_{10} $ in Eqs. \ref{27}.

  Applying the Poisson summation formula, $\ $
  \begin{equation}
\sum_{n=0}^{\infty}\varphi(n)=\sum_{r=-\infty}^{\infty}\int_{0}^{\infty}%
\exp(2\pi irn)\varphi(n)dn.\label{28}%
\end{equation}
  to Eq. \ref{24}, we get:
  \begin{align}
R   = &
 -\frac{1}{4\pi\hbar\lambda^{2}} \sum_{\sigma}\int dn \int dp_z
\frac{df(E_{n,\sigma} (p_z))}{d E_{n,\sigma} (p_z)} v_{0} (p_z) p_z
\nonumber\\ & \times %
 \left\{ 1 + 2 \mbox{Re} \sum_{r=1}^{\infty}\exp(2\pi irn) \right\}.\label{29}%
 \end{align}
  So, we see that oscillating terms appear in the expressions for $R $ and other averages over the Fermi surface included in Egs. \ref{22}, \ref{23}. Due to this reason, an oscillating term occurs in the resulting formula for the
renormalized longitudinal velocity $v_{\sigma}(p_z). $ 
This oscillating term originates from the Fermi-liquid interactions between charge carriers, and it appears only when the FL coefficients $\varphi_{10} $ and $\psi_{10} $ are taken into account, that is, beyond the contact approximation for the Coulomb interaction.

We get the following results for the oscillating parts of $R,R':$
  \be 
  \tilde R = 2 N \left(\frac{B}{F}\right) \Delta, \qquad R' = 2 N \left(\frac{B}{F}\right) \Delta^s , \label{30}
   \ee
      where $N$ is the electron density, and the functions $ \Delta$ and $ \Delta^s $ have the form:
    \begin{align}
  \Delta = & \sum_{r=1}^\infty \frac{(-1)^r}{\pi r} D (r) \sin 
\left(2\pi r \frac{F}{B} \right) \cos \left(\pi r \frac{\omega_0^*}{\omega^*} \right)
 \nn\\ & \times
 J_0 \left(4\pi r 
\frac{t}{\hbar \omega^*}\right), \label{31}
  \\
 \Delta^s = &\sum_{r=1}^\infty \frac{(-1)^r}{\pi r} D (r)  \cos \left
(2\pi r \frac{F}{B} \right)  \sin \left(\pi r \frac{\omega_0^*}{\omega^*} \right)
 \nn\\ & \times
 J_0 \left(4\pi r \frac{t}{\hbar \omega^*}\right).\label{32}
   \end{align}
   Here, $F = cA_0/2\pi\hbar e,\ A_0 $ is the FS cross-sectional area at $ p_z = \pm p_0/2;$ and the cyclotron quantum $\hbar\omega*$ and spin-splitting energy $\hbar\omega_0^* $ are renormalized according to Eqs. \ref{14}, \ref{17}. The damping factor $ D (r) $ describes the effects of the 
temperature and electron scattering   on the magnetic quantum oscillations, 
and $ J_0 (x) $ is the Bessel function. 
The simplest and well known approximation for $D(r)$ equates it to the product $ R_T (r) R_\tau (r) $ where $ R_T (r) = rx /\sinh (rx)\ (x = 2 \pi^2 k_B T/\hbar\omega^* )$ is the temperature factor and $ R_\tau {\bf( r)} = \exp [-\pi r /\omega^* \tau] $ is the Dingle factor describing the effects of electrons scattering characterized by the scattering time $\tau. $ 
The temperature factor appears in the Eqs. \ref{31},\ref{32} as a result of standard calculations repeatedly described in the relevant works starting from the LK paper \cite{4}. The Dingle factor cannot be straightforwardly computed starting from the expessions like  \ref{25}, \ref{26}. This term is phenomenologically included in the Eqs \ref{31},\ref{32} in the same way as in the Shenberg's book \cite{1}. Under low temperatures required to observe magnetic quantum oscillations, the value of $\tau $ is mostly determined by the impurity scattering.  

Using the microscopic many-body perturbation theory it was shown that in this case the Dingle term retains its form, and the corresponding relaxation time could be expressed in terms of the electron self-energy part $\Sigma $ arising due to the presence of  impurities, namely: $ \tau^{-1} = 2 \mbox{Im} \Sigma/\hbar. $ In strong magnetic fields the self-energy $\Sigma $ gains an oscillating term which describes quantum oscillations of this quantity \cite{23,26}. So, the scattering time becomes dependent of the magnetic field $\bf B.$ A thorough analysis carried out in the earlier works of Champel and Mineev \cite{26} and Grigoriev \cite{23} shows that the oscillating correction to the scattering time could be neglected when the FS of a Q2D metal is noticeably warped $(4\pi t> \hbar \omega^*).$ In such cases one may treat $ \tau $ as a phenomenological constant. However, when the FS is very close to a pure cylinder $(4\pi t \ll \hbar \omega^*)$ the scattering time oscillations must be taken into consideration in studies of the de Haas-van Alphen effect. These oscillations may bring some changes in both shape and magnitude of the magnetization oscillations but we do not discuss the issue in the present work. In further analysis we assume that $ 4\pi t > \hbar\omega^*.$

One may notice that  the oscillating 
function $ \Delta $ has exactly the same form as that describing the 
magnetization  oscillations in Q2D metals when the Fermi-liquid effects 
are omitted from the consideration (see   e.g. Refs. \cite{23,24}). Also, 
the Fermi-liquid terms included in the expression (\ref{27}) exhibit  oscillations in the strong magnetic field. For instance, applying the Poisson summation formula   to the expressions (\ref{27}) we can convert these expressions to the form: $ A =\overline a_1 (1+ \delta),\    A' = \overline a_1 \delta^s $ where the oscillating functions $ \delta $ and $ \delta^s $ are
   \begin{align}
  \delta = &\sum_{r=1}^\infty (-1)^r D (r) \cos\left(2\pi r \frac{F}{B}\right)
 \cos\left(\pi r \frac{\omega_0^*}{\omega^*}\right ) S \left(\frac{4\pi rt}{\hbar \omega^*} \right)
  \nn\\ 
-&\sum_{r=1}^\infty (-1)^r D (r) \sin \left(2\pi r \frac{F}{B}\right )
 \cos\left(\pi r \frac{\omega_0^*}{\omega^*}\right ) Q \left(\frac{4\pi rt}{\hbar \omega^*} \right) ,\label{33}
   \end{align}
 \begin{align}
 \delta^s=&\sum_{r=1}^\infty (-1)^r D (r) \sin \left(2\pi r \frac{F}{B}\right)
 \sin\left(\pi r   \frac{\omega_0^*}{\omega^*}\right ) S 
\left(\frac{4\pi rt}{\hbar \omega^*} \right)
   \nn\\
  + &\sum_{r=1}^\infty (-1)^r D (r) \cos \left(2\pi r \frac{F}{B}\right ) 
\sin \left(\pi r
\frac{\omega_0^*}{\omega^*}\right ) Q \left(\frac{4\pi rt}
{\hbar \omega^*} \right) . \label{34}
  \end{align}
  The factors $S $ and $ Q $ entered in Eqs. \ref{33} and \ref{34} are  expressed  in the series of the Bessel functions:
  \bea 
  &&S (x) = J_0 (x) + \frac{3}{2 \pi^2} \sum_{m=1}^\infty \frac{(-1)^m}{m^2} 
J_{2m} (x), \label{35}\\
 && Q(x) = \frac{6}{\pi^2} \sum_{m=0}^\infty \frac{(-1)^m}{(2m + 1)^2} 
J_{2m +1} (x). \label{36}
   \eea
   The expressions for $ B, B' $ are similar to those for $ A, A' $ and we may 
get to former by replacing the factor $\overline a_1$ by another constant $\overline b_1. $ 
The oscillating function $ \delta $ behaves like the function describing quantum oscillations of the charge carriers density of states (DOS) on the FS of a Q2D metal (see Appendix). 
 As for the parameters $\overline a_1,\overline b_1$ we can define $\overline a_1 = - \nu_0 (0) p_0^2 \varphi_{10}/3 $  and $ \overline b_1 $ is similarly defined, namely: $\overline b_1 = -\nu_0(0) p_0^2 \psi_{10}/3. $ We remark that the parameter $\overline a_1 $ differs from $a_1$ which enters the expression for the cyclotron mass (see Eq. \ref{17}). This reflects the anisotropy of electron properties in Q2D conductors.

\section{iv. Quantum oscillations in the magnetization}

To compute the longitudinal magnetization $M_{||},$ we start from the standard
expression:
 \begin{equation}
M_{||}(B,T,\mu)\equiv M_{z}(B,T,\mu)=-\left(  \frac{\partial\Omega}{\partial
B}\right)  _{T,\mu}\label{37}%
  \end{equation}
  Here, the magnetization depends on the temperature $T$ and on the chemical
potential of the charge carriers $\mu,$ and $\mathbf{H}$ is the external
magnetic field related to the field $\mathbf{B}$ inside the metal as
$\mathbf{B=H}+4\pi\mathbf{M.}$ When the magnetic field is directed along a
symmetry axis of a high order we may assume that the fields $\mathbf{B}$ and
$\mathbf{H}$ are parallel. One may neglect the difference between $\mathbf{B}
$ and $\mathbf{H}$ when the magnetization is weak. Otherwise the
uniform magnetic state becomes unstable, and the Condon diamagnetic domains
form, with the alternating signs of the longitudinal magnetization \cite{28}.
We will discuss this possibility later. Now, we assume $H $ by $B$ in
the Eq. \ref{37}. To incorporate the effects of electron interactions  we
assume, in the spirit of the FL theory, that the thermodynamic
potential $\Omega$ has the same form as for noninteracting
quasiparticles, but with the quasiparticle energies fully renormalized by
their interaction:
  \be
\Omega=-k_B T\sum_{\nu}\ln\left\{  1+\exp\left[  \frac{\mu-E_{\nu}}{k_B T
}\right]  \right\}  .\label{38}%
  \ee
   In this expression $E_{\nu}$ is the quasiparticle energy including the
correction arising due to the FL interactions, and $k_B $ is the Boltzmann's constant.

Accordingly, we rewrite Eq. \ref{38} as follows:
 \be
\Omega=-\frac{k_B T}{4\pi^{2}\hbar\lambda^{2}}\sum_{n,\sigma}\int\ln\left\{
1+\exp\left[  \frac{\mu-E_{n,\sigma}(p_z)}{k_B T}\right]  \right\}
dp_z \label{39} %
  \ee
  where $\lambda^{2}=\hbar c/eB$ is the squared magnetic length.
Performing integration by parts, Eq. \ref{39} becomes:
  \begin{equation}
\Omega=-\frac{1}{4\pi^{2}\hbar\lambda^{2}}\sum_{n,\sigma}\int f(E_{n,\sigma
}(p_z))v_{\sigma}(p_z)p_z dp_z. \label{40}%
     \end{equation}
 Applying the Poisson summation formula, we get:
  \begin{align}
 \Omega  = - &\frac{1}{4\pi^{2}\hbar^{3}\lambda^{2}}\sum_{\sigma}\int dn\int
dp_zf(E_{n,\sigma}(p_z))v_{\sigma}(p_z) p_z 
   \nn\\ \times &
\left\{  1+2\mbox{Re}\sum_{r=1}^{\infty}\exp[2\pi irn]\right\}
. \label{41}%
  \end{align}
  So we see that the expression for the thermodynamic potential includes two oscillating terms. One originates from the oscillating part of
$v_{\sigma}(p_z).$ The second term inside the braces in the Eq. \ref{41} gives
another oscillating contribution.

The effects of temperature and spin splitting on the magnetic oscillations are
already accounted for in the Eq. \ref{41}. Assuming $4\pi t > \hbar\omega, $ we take into account the effect of
electron scattering  adding an imaginary part $i\hbar/2\tau$ to the electron
energies  \cite{1}. After standard
manipulations, we obtain the following expression for the oscillating part of
the longitudinal magnetization:
 \begin{align}
   \Delta & M_{||} =- 2 N \beta \frac{\omega_0^*}{\omega^*}(1 - 3a_1^*)
   \nn \\   \times &  
\frac{\Delta - 3(a_1^* + b_1^*)\delta + 3b_1^*(\Delta \delta 
- \Delta^s\delta^s) - 9a_1^* b_1^*(\delta^2 - \delta^{s2})}{1 +3 (a_1^* + b_1^*) \delta + 9a_1^* b_1^*(\delta^2 - \delta^{s2})} . \label{42}
   \end{align}
  where $a_1^* =\overline a_1/(1 + 3\overline a_1);\ b_1^* =\overline b_1/(1 + 3\overline b_1).$ This is the main result of the present work. It shows that the Fermi-liquid interactions may bring significant changes in the de Haas-van Alphen oscillations. Below, we analyze these changes.  If we may neglect the oscillating corrections proportional to $ a_1^*, b_1^*, $ then our result for 
$\Delta M_{||}$ reduces
to the usual LK form with some renormalizations arising from the quasiparticle
interactions. The cyclotron mass $m_{\perp}$ differs from the bare Fermi
liquid cyclotron mass before the quasiparticle interaction is taken into
account (cf. Eq. \ref{17}), and $\hbar\omega_0^*$ includes the extra factor
$(1+b_{0})^{-1}$. Also, the factor $(1-3a_{1}^{\ast})$ modifies the magnetic oscillations magnitudes. As for the oscillations frequencies, they remain unchanged by the FL interactions, as expected.

\section{v. Discussion}

Comparing our result (\ref{42}) with the corresponding result reported by
Wasserman and Springfield \cite{6}, we see that these results agree with each
other. A seeming difference in the expressions for the oscillations
frequencies arises due to the fact that in Ref. \cite{6} the frequencies are
expressed in terms of the chemical potential of electrons $\mu$ instead of the
cross-sectional areas of the Fermi surface. It is  worth reiterating that
\textquotedblleft unrenormalized\textquotedblright\ mass in the FL theory is
already renormalized (sometimes strongly) from fully noninteracting (or
density functional - calculated) mass. Again, we remark that the present analysis was carried out assuming noticeable/significand FS warping $(4\pi t> \hbar\omega^*),$ so, we may neglect the magnetic field dependence of the electrons scattering $\tau $ treating the latter a constant phenomenological parameter. This results in a simple form of the Dingle damping factor $ R_\tau (r) $ describing the effects of electrons impurity scattering.

The LK form of the expression for the longitudinal magnetization is suitable
to describe de Haas-van Alphen oscillations in conventional three dimensional
metals within the whole range of temperatures. However, this is not true for
quasi-two-dimensional conductors. The Fermi surface of such a conductor is
nearly cylindrical in shape, therefore the oscillating term in the denominator of Eq. \ref{42} significantly increases. The oscillations of the denominator of Eq. \ref{42} occur due to the functions $ \delta $ and $ \delta^s. $ These functions are presented in the Fig. 1, and we see that at  low temperatures and weak scattering
$\ln (\hbar\omega^*/k_B T^*) > t/\hbar\omega^*$ $(T^{\ast}=T +T_{D},\ T_{D}=\hbar/2\pi k_{B}\tau$ is the Dingle temperature)  the peak values may be of the order of $1$, especially for a rather weakly warped FS $(t/\hbar\omega^* \sim 0.1\div0.5).$
 So, quantum oscillations in the magnetization 
in the electron Fermi-liquid in quasi-two-dimensional metals may have more complicated structure than those in the electron gas described by the
LK formula \cite{4}. The effect of the Fermi-liquid interactions on these oscillations depends on the values of the Fermi-liquid parameters $ a_1^*, b_1^*$ and on the damping factor $D(r) = R_T (r) R_\tau (r) $ included in the expressions for the oscillating functions. The FS shape determined by the ratio $ t/\hbar\omega $ is important as well.

\begin{figure}[t]
\begin{center}
\includegraphics[width=4.4cm,height=8.6cm,angle=-90]{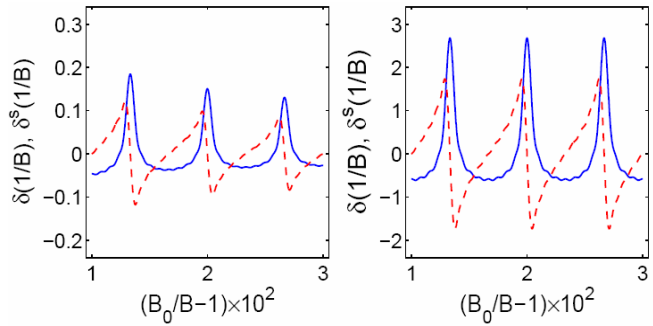}
\end{center}
\caption{The magnetic field dependencies of the functions $\delta $ (solid lines) and $\delta^s $ (dashed lines). The curves are plotted at $B_0 = 10T,\ F/B_0 = 300,\ 2\pi^2\theta^*/\hbar\omega^* = 0.5 $ for $ t/\hbar\omega^* = 2 $ (left panel) and $ t/\hbar\omega^* = 0.3 $ (right panel). }%
\label{rateI}%
\end{figure}

The most favorable conditions for the changes in the magnetization oscillations  to be revealed occur when the oscillating terms in the denominator of the Eq. (42) may take on values of the order of unity at the peaks of oscillations. 
We may estimate the peak values $ \delta_m $ and $ \delta_m^s $ of the 
functions included into the above denominator using the Euler-Macloren formula.  The estimations depend on the shape of the FS of the Q2D conductor. 
When the FS is significantly crimped $(t\gg \hbar \omega)$ we obtain: 
$ \delta_m, \delta_m^s \sim (\hbar \omega /t)^{1/2} (k_B T^*)^{-1/2}.$  So, we may expect the Fermi-liquid interaction to be distinctly manifested in the 
magnetization oscillations when $|a_1^*| (\hbar \omega/ t)^{1/2}(k_B T^*)^{-1/2} \sim 1 $ or $|b_1^*| (\hbar \omega/ t)^{1/2} (k_B T^*)^{-1/2} \sim 1 $ or both. In all probability the Fermi-liquid parameters are small in magnitude 
$(|a_1^*|,|b_1^*| \ll 1). $ Nevertheless, the changes in the de Haas-van Alphen oscillations arising due to the Fermi-liquid effects may occur at $k_B T^* \ll 1. $ It is worthwhile to remark 
that due to the character of the electron spectra, the Q2D conductors provide 
better opportunities for observations of the Fermi-liquid effects in the de 
Haas-van Alphen oscillations than conventional 3D metals. In the latter the 
peak values of the oscillating functions $ \delta, \delta^s $ have the order 
$(\hbar \omega/\mu)^{1/2} (k_B T^*)^{-1/2}. $ Typical values of the transfer 
integral $ t $ are much smaller than those of the chemical potential $ \mu, $ 
therefore significantly smaller values of $ k_B T^* $ and/or greater values 
of the parameters $ a_1^*, b_1^* $ are required for the Fermi-liquid effects to be revealed in 3D metals.

\begin{figure}[t]
\begin{center}
\includegraphics[width=8.8cm,height=8.6cm,angle=-90]{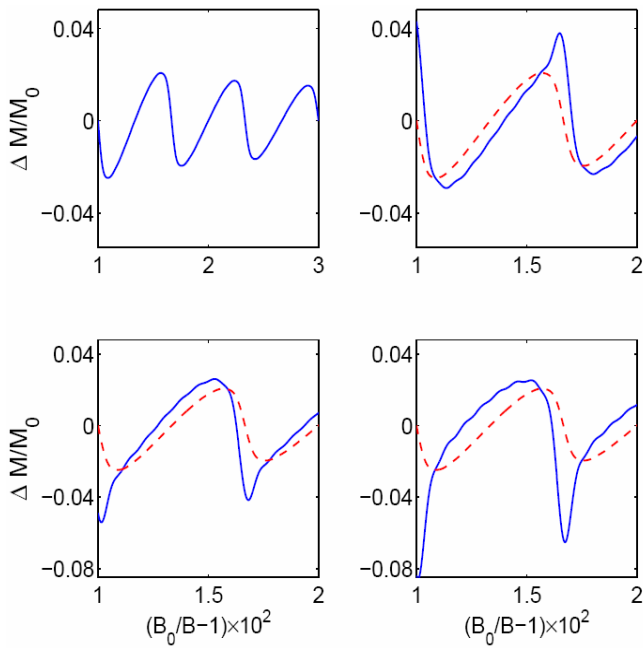}
\caption{
The effect of the Fermi-liquid interactions on the de Haas-van Alphen 
oscillations in a Q2D metal at $ t/\hbar \omega^* = 2. $ The curves are plotted using Eq. \ref{42}, $ M_0 = 2 N\beta. $ Calculations are carried out for $ a_1^* = b_1^* = 0 $ (top left panel) $ a_1^* = b_1^* = 0.02 $ (top right panel), $ a_1^* = b_1^* =-0.02;  -0.04 $ (bottom left and right panel, respectively). The remaining parameters are the same as used in the figure 1. The dashed lines in the top right panel and in the bottom panels represent oscillations in the system of noninteracting quasiparticles.}
 \label{rateI}
\end{center}
\end{figure}

\begin{figure}[t]
\begin{center}
\includegraphics[width=8.8cm,height=8.6cm,angle=-90]{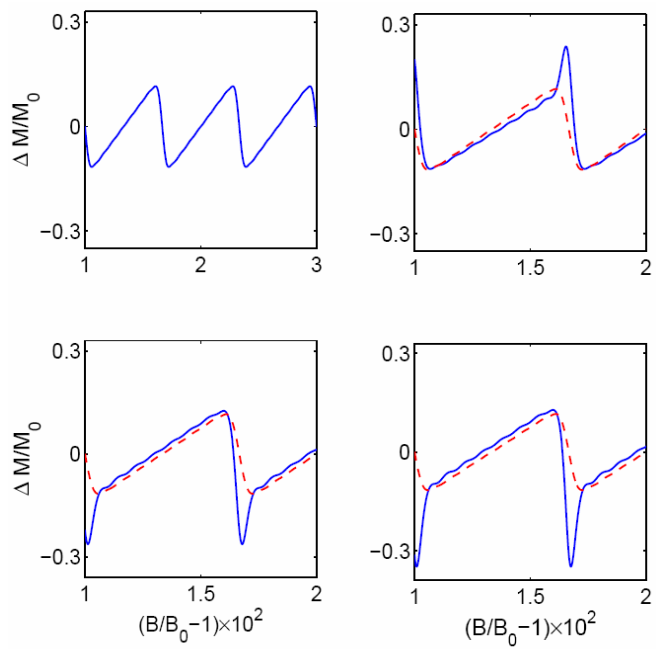}
 \caption{ 
The effect of the Fermi-liquid interactions on the de Haas-van 
Alphen oscillations in a Q2D metal at $ t/\hbar \omega^*=0.3. $ The curves 
are plotted using Eq. \ref{42} at $ a_1^* = b_1^* = 0 $ (top left panel), $ a_1^* = b_1^* = 0.02 $ (top right panel), $ a_1^* = b_1^* =- 0.02 $ (bottom left panel), and $ a_1^* = b_1^* = -0.04 $ (bottom right panel). The remaining parameters are the same as in the figure 1. Dashed lines represent the oscillations in the gas of charge carriers.}
 \label{rateI}
\end{center}
\end{figure}

Due to the special character of the electron spectra 
in the Q2D metals, the variations in the magnetization oscillations may be 
noticeable at reasonably small values of the Fermi-liquid constants. In the 
Fig. 2 we compare the oscillations arising in a gas of the charge carriers 
(top left panel) with those influenced by the Fermi-liquid interactions 
between them. All curves included in this figure are plotted within the limit 
$ t > \hbar \omega. $ We see that both magnitude and shape of the 
oscillations noticeably vary due to the Fermi-liquid effects.     
When $t/\hbar\omega^* \sim 0.1\div0.5 $ the FS shape is closer 
to a perfect unwarped cylinder, the magnetization oscillations accept the well 
known sawtoothed shape, shown in the Fig. 3. Again, when the Fermi-liquid 
interaction produced terms are included into the expression for $\Delta M, $ 
this bring some changes in the magnitude and shape of the oscillations. These changes are more significant when $(a_1^* + b_1^*) < 0. $ 
 
   \begin{figure}[t]
\begin{center} \includegraphics[width=4.4cm,height=8.6cm,angle=-90]{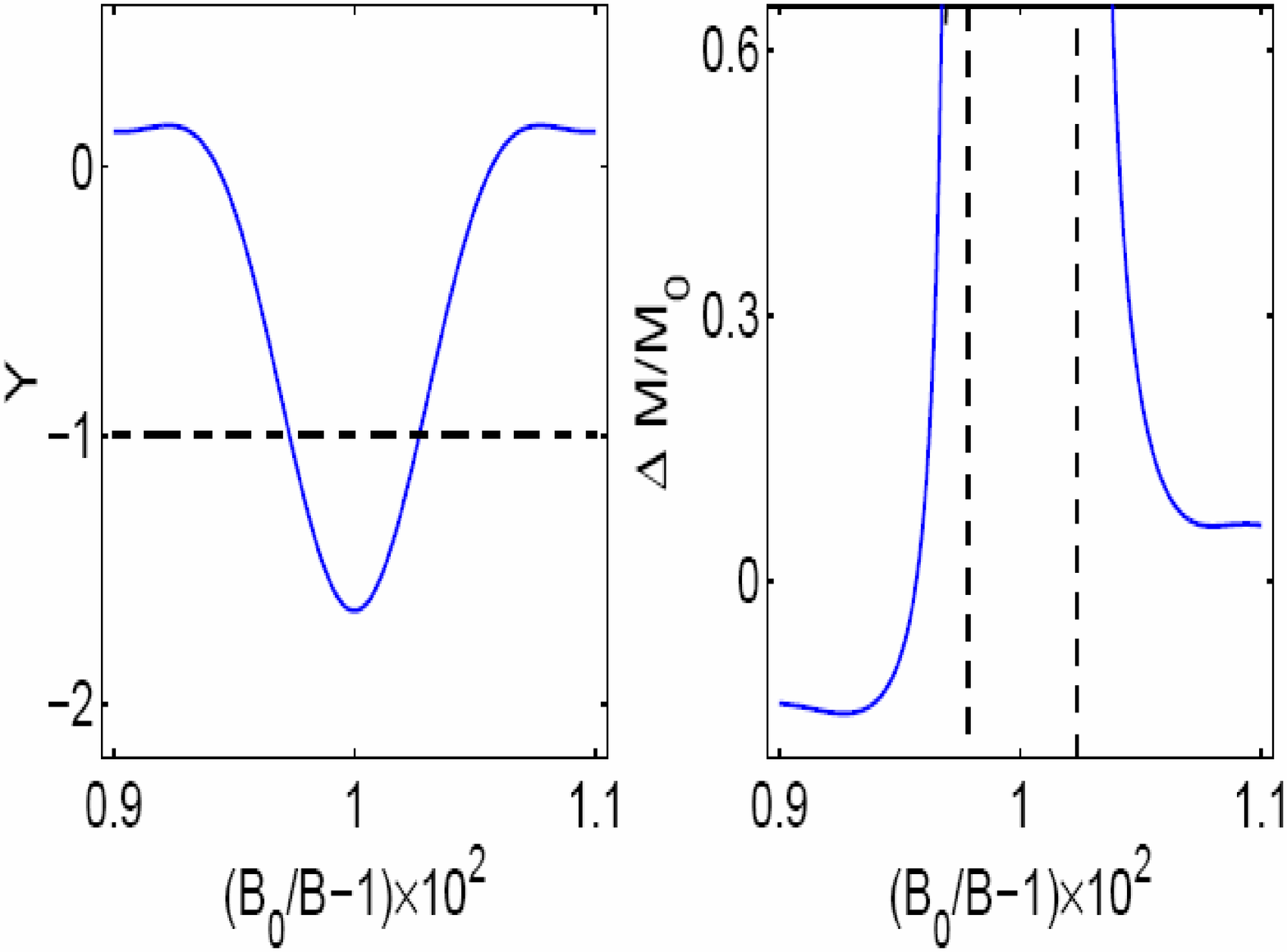}
\caption{
The plot of the  function $ Y = (a_1^* + b_1^*)\delta+ a_1^* b_1^* (\delta^2
 - \delta^{s2}) $ near the peak of the DOS quantum oscillations at 
$ t/\hbar \omega^* =0.3 $ which illustrates that the denominator in the Eq. \ref{42} may become zero in the vicinities of these peaks (left panel). The divergencies in the oscillating part of magnetization described by Eq. \ref{42} indicating the magnetic instability caused by the Fermi-liquid effects (right panel). The curves are plotted assuming $ a_1^* = b_1^* = -0.04. $ The remaining parameters have the same values as used in the figure 1.}
\label{rateI} 
\end{center}
\end{figure}

The most important manifestation of the Fermi-liquid effects occurs in very clean conductors at low temperatures when 
$ T^*$ is reduced so much that $(a_1^* + b_1^*)\delta_m $ is greater than $1. $ Then the denominator of the Eq. \ref{42} becomes zero at some points near the peaks of the DOS oscillations provided that $(a_1^* + b_1^*) < 0. $ This is illustrated in the Fig. 4. Correspondingly, $ \Delta M $ diverges at these points which indicates the magnetic instability of the system.

This means that the condition for
the uniform magnetization of the electron liquid is violated near the
oscillations maxima, and the diamagnetic domains could emerge. It is known
that both crystal anisotropy and demagnetization effects originating from the
shape of the metal sample, could modify the relation between $\mathbf{B}$ and
$\mathbf{H}$ and cause magnetic instability which results in the occurence of
the diamagnetic domains \cite{28}. Our result demonstrates that the
interactions of conducting electrons also may play an important role in 
magnetic phase transitions.
 In principle, such magnetic instabilities 
may appear in 3D metals as well which was shown earlier analyzing quantum 
oscillations in the longitudinal magnetic susceptibility of the isotropic 
electron liquid (see Refs. \cite{18,29}). However, we may hardly expect these 
transitions to appear in conventional metals for the requirements on the 
temperature and intensity of scattering processes are very strict. Estimations 
made in the earlier works \cite{20,29} show that the temperature (including the Dingle correction) must be about $ 10 mK $ 
or less for these diamagnetic phase transitions to emerge in conventional 
metals. On the contrary, the special (nearly cylindrical) shape of the FS in Q2D conductors gives grounds to expect the above transitions to appear in realistic experiments.

 To summarize, in the present work we theoretically analyzed possible
manifestations of the FL interactions (that is, residual interactions of excited quasiparticles) in the de Haas-van Alphen oscillations in Q2D conductors. The same approach can be easily applied for a metal with the crystalline lattice of arbitrary symmtery, using the appropriate basis for expanding the FL functions. So, the phenomenological Fermi-liquid theory becomes more realistic and suitable to analyze effects of electron interactions in actual metals.

We showed that the residual quasiparticle interactions affect all damping
factors inserted in the LK formula through the renormalization of the
cyclotron mass. The spin splitting is renormalized as well, in a manner
similar to the so-called Stoner enhancement. The frequency of the oscillations remains unchanged for it is determined with the main geometrical characteristics of the Fermi surface, which probably are not affected by electron-electron interactions. However, the shape and magnitude of the oscillations are affected due to the FL effects, and their changes may be noticeable. Also, the obtained results indicate that (under the relevant conditions) the electron
interactions may break down the magnetic stability of the material creating an
opportunity for the diamagnetic phase transition. The discussed effects may be
available for observations in realistic experiments bringing extra
informations concerning electronic properties of quasi-two-dimensional metals.

Finally, we want to emphasize once again that the renormalizations, most
importantly, mass renormalization, are \textit{in addition} to what is
conventionally called \textquotedblleft mass renormalizatin\textquotedblright,
namely, renormalization of the specific heat coefficient compared to band
structure calculations. It is usually implicitely assumed that the weighted
average of the de Haas-van Alphen mass renoramization is exactly equal to the
specific heat renormalization, i.e., the FL effects are small. In many cases
this is a good approximation, but one can never exclude a possibility that in
some materials these two masses may be different, namely, the de Haas-van
Alphen mass may be larger. A curious example when one would have needed to
exercise caution, but did not, is given by Ref. \cite{30}, where quantum
oscillations in a highly uncoventional metal, Na$_{x}$CoO$_{2},$ were
measured, and it was taken for granted that the large mass renormalization
found in the experiment should be fully accounted for in specific heat. Based
on this assumption, a natural and straightforward interpretation of the data
was abandoned and  counterintuitive explanation, requiring some unverified assumptions was accepted. It is possible that the results reported in the Ref. \cite{30} give a case where additional mass renormalization discussed in this paper is significant. Hopefully, at some point we will see a careful and accurate experimental study on various materials that would compare the de Haas-van Alphen masses with the thermodynamic masses and give us a quantitative answer on how different may the two be in real life.

\section{appendix}

Here, we analyze the expressions for the oscillating functions $ \delta,\delta^s$ within the limits of significant $(t \gg \hbar\omega)$  FS warping. We may use the standard asymptotics for the Bessel functions , at $ x\gg 1, $ namely:
  \be 
J_k (x) \approx \sqrt{\frac{2}{\pi x}} \cos \left[x - \frac{\pi k}{2}- \frac{\pi}{4}\right].\label{43}
  \ee
 Substituting these aproximation into Eqs. \ref{34}, \ref{35} we obtain:
   \begin{align}
 & S(x) = \sqrt{\frac{2}{\pi x}} \cos \left[x- \frac{\pi}{4}\right] \left\{1+ \frac{3}{\pi^2} \sum_{m=1}^\infty \frac{1}{m^2} \right\},\label{44}
  \\
& Q(x) = \frac{6}{\pi^2} \sqrt{\frac{2}{\pi x}} \sin \left[x - \frac{\pi}{4}\right] \sum_{m=0}^\infty \frac{1}{(2m +1)^2}, \label{45}
  \end{align}
   where 
  \be
 \sum_{m=1}^\infty \frac{1}{m^2} = \frac{\pi^2}{6}, \qquad
\sum_{m=0}^\infty \frac{1}{(2m+1)^2} = \frac{\pi^2}{8}. \label{46}
  \ee
  So, we have:
  \begin{align}
 & S(x) = \frac{5}{4} \sqrt{\frac{2}{\pi x}} \cos\left[x - \frac{\pi}{4}\right], 
\label{47}\\
& Q(x) = \frac{3}{4} \sqrt{\frac{2}{\pi x}} \sin\left [x - \frac{\pi}{4}\right].
  \label{48}
  \end{align} \begin{widetext}
 Using these results we may write the following expressions for $ \delta,\delta^s $ at $t \gg \hbar\omega: $ 
   \begin{align}
 \delta =& \frac{5}{4} \left(\frac{\hbar\omega^*}{2\pi^2t}\right)^{1/2} 
\sum_{r=1}^\infty \frac{(-1)^r}{\sqrt r} D(r)
 \cos\left[2\pi r\frac{F}{B}\right] 
\cos\left[\frac{4\pi rt}{\hbar\omega^*} - \frac{\pi}{4}\right] 
\cos\left[\pi r \frac{\omega_0^*}{\omega^*}\right]
  \nn\\ &
-\frac{3}{4} \left(\frac{\hbar\omega^*}{2\pi^2t}\right)^{1/2} 
\sum_{r=1}^\infty \frac{(-1)^r}{\sqrt r} D(r)
 \sin\left[2\pi r\frac{F}{B}\right] 
\sin\left[\frac{4\pi rt}{\hbar\omega^*} - \frac{\pi}{4}\right] 
\sin\left[\pi r \frac{\omega_0^*}{\omega^*}\right] . \label{50}
  \end{align}
  Or:
  \be
\delta =\left(\frac{\hbar\omega^*}{2\pi^2t}\right)^{1/2} \sum_{r=1}^\infty
\frac{(-1)^r}{\sqrt r} D(r) \cos\left[\pi r \frac{\omega_0^*}{\omega^*}\right]
\left\{\cos\left[2\pi r\frac{F_{\max}}{B} - \frac{\pi}{4} \right] + 
\frac{1}{4}\cos\left[2\pi r\frac{F_{\min}}{B} +\frac{\pi}{4}\right] \right\}. \label{51}
 \ee
 Likewise, we obtain for $ \delta^s:$
   \be
\delta =\left(\frac{\hbar\omega^*}{2\pi^2t}\right)^{1/2} \sum_{r=1}^\infty
\frac{(-1)^r}{\sqrt r} D(r) \cos\left[\pi r \frac{\omega_0^*}{\omega^*}\right]
\left\{\sin\left[2\pi r\frac{F_{\max}}{B} - \frac{\pi}{4} \right] + \frac{1}{4}
\sin\left[2\pi r\frac{F_{\min}}{B} +\frac{\pi}{4}\right] \right\}. \label{52}
   \ee
  Here, $ F_{\max}, F_{\min}$ correspond to the maximum and minimum cross-sectional areas of the FS, respectively.
\end{widetext}


\section{ Acknowledgments}

Author thanks G. Kotliar and I. Mazin for helpful discussions, and G. M. Zimbovsky for help with the manuscript. This work was supported by DoD grant W911NF-06-1-0519 and NSF-DMR-PREM 0353730.

\end{document}